\documentclass[superscriptaddress,showpacs,preprintnumbers,showkeys,nofootinbib]{revtex4}

\usepackage{amsmath,amssymb}
\usepackage{graphicx}

\begin{document}
\preprint{DO-TH 12/18}

\title{Geometrical CP violation in multi-Higgs models}

\author{Ivo de Medeiros Varzielas}
\email{ivo.de@udo.edu}
\affiliation{Fakult\"{a}t f\"{u}r Physik, Technische Universit\"{a}t Dortmund D-44221 Dortmund, Germany}

\keywords{Discrete non-Abelian symmetries, multi-Higgs models, CP violation}

\pacs{11.30.Hv, 12.60.Fr}


\begin{abstract}
We introduce several methods to obtain calculable phases with geometrical values that are independent of arbitrary parameters in the scalar potential. These phases depend on the number of scalars and on the order of the discrete non-Abelian group considered. Using these methods we present new geometrical CP violation candidates with vacuum expectation values that must violate CP (the transformation that would make them CP conserving is not a symmetry of the potential). We also extend to non-renormalisable potentials the proof that more than two scalars are needed to obtain these geometrical CP violation candidates.
\end{abstract}


\maketitle

\section{Introduction \label{Intro}}

The groups $\Delta(3 n^2)$ and $\Delta(6 n^2)$ \cite{FFK, Bovier:1980gc, Luhn:2007uq, Escobar:2008vc, Ishimori:2010au, Merle:2011vy, Grimus:2011fk} have been increasingly used in phenomenological studies, with the use of $\Delta(27)$ to obtain patterns of leptonic mixing (e.g. \cite{deMedeirosVarzielas:2006fc}) contributing to renewed interest in them. Historically, $\Delta(27)$ had been applied to hadron physics \cite{FFK} and to obtain candidates of geometrical CP violation (GCPV) \cite{Branco:1983tn}: complex vacuum expectation values (VEVs) with calculable phases $\pm2 \pi/3$ determined entirely by the symmetry of the scalar potential, within a framework of spontaneous CP violation \cite{Lee:1973iz,Branco:1979pv}. The predictive power of GCPV is quite appealing and motivated further study - it was shown that $\Delta(54)$ and other $\Delta(3 n^2)$ and $\Delta(6 n^2)$ groups can lead to the same calculable phases $\pm 2 \pi/3$, with prospects of viable fermion mass structures \cite{deMedeirosVarzielas:2011zw}.
It was also shown recently that these GCPV solutions can be preserved up to arbitrary high orders \cite{Varzielas:2012nn} beyond the renormalisable level considered in \cite{Branco:1983tn,deMedeirosVarzielas:2011zw}.
Here we introduce methods to obtain other calculable phases that could lead to GCPV candidate VEVs with powers of $\eta_n \equiv e^{\text{i} 2 \pi / n}$ for any $n$.
We consider implicitly multi-Higgs doublet extensions of the Standard Model (SM): each scalar $H_i$ is one $SU(2)$ doublet out of a total of $N$ doublets. This needs not be the case though and it will be apparent that the results obtained apply in general to scalar potentials as long as some mechanism enforces each invariant to have an equal number of conjugate and unconjugated scalar components.

\section{Calculable phases}

Following the discussion in \cite{Varzielas:2012nn}, for $N=3$ scalars in a single irreducible representation $H_i$ of $\Delta(3 n^2)$ or $\Delta(6 n^2)$ groups, the nature of the group forces terms in the potential to be invariant under cyclic permutations of the $H_i$.
It is helpful to use the parametrisation $\langle H_i \rangle = v e^{\text{i} \alpha_i}$ (note the VEVs have the same magnitude $v$). In a given invariant we denote the overall phase of each of the cyclic permutations as $A_i$, a linear combinations of the individual phases $\alpha_i$. Specific examples of these phase dependences where previously discussed
in \cite{deMedeirosVarzielas:2011zw, Varzielas:2012nn}.
For $N$ scalars the most general cyclic phase dependences can be expressed by $A_1 = \sum a_i \alpha_i$ and cyclic permutations $A_2 = \sum a_{i-1} \alpha_i$ up to $A_N =\sum a_{i-N+1} \alpha_i$, with any $a_i$ with non-positive index corresponding to that index plus $N$ (e.g. $a_0=a_N$, $a_{-1} =a_{N-1}$).
We can then obtain $\sum A_i = \left( \sum a_i \right) \left( \sum \alpha_i \right)$.

Specific constraints on the type of invariants allowed reflect on $\sum a_i$ and can have important implications. Consider $H_1 H_2 H_3$ and $H_1^3+c.p.$, where $c.p.$ denotes all possible cyclic permutations. They both verify $\sum a_i = 3$ which can be a consequence of the discrete non-Abelian symmetry considered. If these examples are present in the potential with a positive (or negative) coefficient, $H_1 H_2 H_3$ is minimised if its phase dependence $\alpha_1 + \alpha_2 + \alpha_3$ is equal to $\pi$ (or $0$), whereas $H_1^3+c.p.$ is minimised when the same conditions are verified simultaneously for its three $A_i=3 \alpha_i$. Therefore if these terms appear together and both have negative coefficients, they can align $H_i$ to be real (apart from an overall phase).
The analysis of other such terms can be of great interest, and specific orders can be selected by using an additional commuting $C_M$ cyclic group. To remove $H_1 H_2 H_3$, $H_1^3+c.p.$ one could use e.g. $C_6$ and preserve $H_1^6+c.p.$ and $(H_1 H_2 H_3)^2$. A more detailed analysis of these cases is beyond the scope of the present work.

In the cases which we will be considering in detail, the constraint $\sum a_i =0$ applies and so $\sum A_i=0$ (for the $A_i$ phase dependences this is apart from adding integer multiples of $2 \pi$). This property can originate from the requirement that each invariant has an equal number of $H$ and $H^\dagger$ components, which in turn can be fulfilled by considering each scalar to be equally charged under a separate group commuting with the non-Abelian discrete symmetry: with an $U(1)$, $\sum a_i =0$ exactly and if one uses instead a $C_M$ cyclic group it provides it as an approximation up to order $M$. We implicitly consider SM doublets but the results obtained apply equally to multi-Higgs doublet extensions of the Standard Model or to any other multi-Higgs scalar potentials that have $\sum a_i =0$.

\begin{figure}
	\centering
		\includegraphics[width=3 cm,keepaspectratio=true]{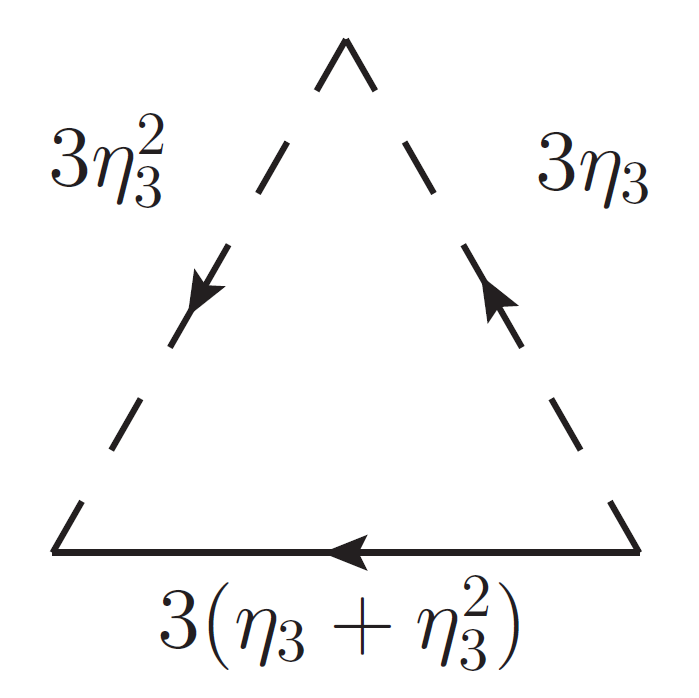}
	\caption{\label{fig:triangle} The equilateral triangle shape associated with minimising solution for $V_3$.}
\end{figure}

With $\sum a_i = 0$ in mind, we consider the source of the known GCPV solutions \cite{Branco:1983tn, deMedeirosVarzielas:2011zw, Varzielas:2012nn}:
\begin{equation}
H_1^2 (H_2 H_3)^\dagger + c.p.= v^4 (e^{\text{i} A_1} + e^{\text{i} A_2} + e^{\text{i} A_3}) \,,
\end{equation}
where the respective $A_i$ are $A_1=2\alpha_1 - \alpha_2 - \alpha_3$, $A_2=-\alpha_1 + 2\alpha_2 - \alpha_3$,  $A_3=-\alpha_1 - \alpha_2 + 2 \alpha_3$. We can identify $a_i$ as the vector $(2,-1,-1)$ representing also the possible cyclic permutations and their hermitian conjugates $h.c.$ - using this notation, the $a_i$ vector identifies the respective invariant. 
In the renormalisable case, the $(2,-1,-1)$ invariant is the only phase dependence in the potential, and the minimum of the scalar potential depends on the quantity
\begin{equation}
V_3 =  (e^{\text{i} A_1} + e^{\text{i} A_2} + e^{\text{i} A_3}) + h.c. \,,
\end{equation}
which is confined to the real axis of the complex plane. If $V_3$ appears in the potential with a negative coefficient, then $V_3$ should be maximised, which corresponds to a contribution of $+6$ for $A_i=0$. If the sign is positive though, $V_3$ should be minimised by the phases - but it is impossible to get $-6$ as $A_i= \pi$ would violate $\sum A_i=0$.
The minimum is $-3$ and corresponds to e.g. $A_i = \pm 2 \pi/3$.
From these solutions we can obtain the known GCPV candidates of the type $(\eta_3^{\mp 1},1,1)$ corresponding to $\alpha_1= \mp 2 \pi / 3$, $\alpha_2=\alpha_3=0$. By definition $\eta_3+\eta_3^2=-1$ and on the complex plane the shape that leads to these VEVs can be seen in figure \ref{fig:triangle} to be an equilateral triangle.

Beyond renormalisable level the only other phase-dependent invariants for $\Delta(27)$ and $\Delta(54)$ are either integer multiples of the renormalisable one - with $a_i$ vector $k(2,-1,-1)$ - or the combinations that saturate $n=3$, with $a_i$ vector $(3,-3,0)$ and its own multiples \cite{Varzielas:2012nn}. The group properties are fundamentally linked to the invariants: the transformations involve powers of $\eta_3$, so the group treats $2$ and $-1$ as equivalent and $k(2,-1,-1)$ is invariant. Similarly $k(3,-3, 0)$ is invariant because multiples of $3$ are equivalent to $0$.

We assume that for any number $N$ of scalars $H_i$, the VEV is of the $(1,(...),1)$ type. We justify this assumption by restricting the symmetry groups considered to contain a cyclic group $C_N$ and generalising the reasoning presented in \cite{Varzielas:2012nn}: we classify invariant combinations according to the number of components featured on a single part of the cyclically invariant combination, from just one component in each part of the sum such as $H_1 H_1^\dagger + c.p.$ to all components such as $(H_1 H_1^\dagger (...) H_N H_N^\dagger)$. Depending on the sign of the combined coefficient of those combinations, it is possible to guarantee a $(1,(...),1)$ VEV as a natural result of large, non fine-tuned regions of the parameter space \cite{Varzielas:2012nn}.
The first class can not vanish unless $v=0$, and favours $(1,0,(...),0)$ or $(1,(...),1)$ respectively for negative or positive combined coefficient. 
The last class with negative coefficient also favours $(1,(...),1)$. The first and last classes tend to dominate so provided their coefficients have the correct signs the $(1,(...),1)$ VEV is naturally favoured.
Regardless of their naturalness, one may wonder what would happen with VEVs containing different magnitudes. A simple case would be a direction like $(x,1,(...),1)$. For $x=0$, most of the phase-dependent invariants identically vanish and the non-vanishing ones will be of type $k(n,-n,0,(...),0)$ (such as $k(3,-3, 0)$) - due to $x=0$, there will be two unphysical $A_i$ phase combinations, and the respective invariants will always be minimised by the remaining physical $A_i=0$ or $A_i=\pi$. Otherwise, for $x \neq 0$ it is the interplay of the coefficients of a multitude of invariants (including phase-independent ones) that determines $x$ - that dependence on paramemters of the potential does not allow calculable phases to arise.
Given that the $(1,(...),1)$ VEV is natural and that other solutions are not interesting for obtaining calculable phases, we consider now only phase-dependent invariants in the case of $(1,(...),1)$.

\subsection{Even number of scalars}

We start with $N=4$ scalars $H_i$ in a single irreducible representation of a group with the cyclic permutation $C_N$, assuming a VEV $\langle H_i \rangle = v e^{\text{i} \alpha_i}$.
For each phase-dependent invariant we have $A_i$ which appear in the potential through
\begin{equation}
V_4 =  (e^{\text{i} A_1} + e^{\text{i} A_2} + e^{\text{i} A_3}+e^{\text{i} A_4}) + h.c. \,,
\end{equation}
and if $V_4$ has a negative coefficient $A_i=0$ maximises $V_4$ to $8$. For positive coefficient, it is possible to reach $-8$ with $A_i= \pi$ not violating $\sum A_i = 0$ (the $A_i$ are phases). $A_i=\pi$ applies equally to minimise $V_N$ for any even $N$.

Interesting invariants can originate from groups with non-commuting $C_4$ generators. For a given representation of size 4, the generator $c_4$ is the cyclic permutation matrix and $d_4$ is a diagonal matrix with unit determinant and entries that are different powers of $\eta_4 \equiv e^{\text{i} 2 \pi/4} = \text{i}$. This semi-direct product is a discrete subgroup of $SU(4)$ and because $d_4$ has different powers of $\eta_4$ the group can be identified as $C_4 \ltimes (C_4 \times C_4 \times C_4)$ with order $256$, i.e. $4 n^3$ with $n=4$ \footnote{This is essentially a generalisation of $\Delta(3 n^2)$ as $C_3 \ltimes (C_3 \times C_3)$ for $n=3$.}. The factor of 4 comes from the four possible cyclic structures $c_4^0$, $c_4^1$, $c_4^2$, $c_4^3$ and the $n^3$ from considering $n$ possible powers of $\eta_n$ chosen independently $3$ times.
Details of the group are beyond the scope of the present work, but from the generators $c_4$ and $d_4$ in the representation chosen for $H_i$ it is possible to verify that the non-renormalisable term $H_1^3 (H_2 H_3 H_4)^\dagger + c.p.$ is invariant. The associated $A_1 = 3 \alpha_1 - \alpha_2 - \alpha_3 - \alpha_4$ and $a_i$ is $(3,-1,-1,-1)$. Considering the respective $V_4$, the solution $A_i = \pi$ is obtained with e.g. single non-zero $\alpha_1 = \pi$.

At higher order there are other phase-dependent invariants: generalising \cite{Varzielas:2012nn} we can identify multiples of the lowest order term with $a_i$ vectors $k(3,-1,-1,-1)$ and the terms that saturate $n=4$: $\left( H_1 H_2^\dagger \right)^4 + c.p.$, its $h.c.$, and the self-conjugate $\left( H_1 H_3^\dagger \right)^4 + c.p.$ with $a_i$ vectors $(4,-4,0,0)$ and $(4,0,-4,0)$ respectively. If they appear with negative coefficients, terms such as these and their multiples that saturate $n$ do not distinguish between phases that are integer multiples of $2 \pi/n$.
In addition to the generalisations from \cite{Varzielas:2012nn}, there are also invariants with $a_i$ vectors such as $(2, 2, -2, -2)$ and the self-conjugate invariant associated with $(2,-2,2,-2)$. If this type of invariants dominate, the $V_4$ minimising solution $A_i = \pi$ is obtained with e.g. single non-zero $\alpha_1 = \pm \pi/2$. But for these higher order invariants to dominate over $(3,-1,-1,-1)$ we need unnatural hierarchies between the respective coefficients.

For any number of scalars $N$ and groups generated by $c_N$ and $d_n$ with $n=N$ the leading order phase-dependent invariant (with associated $a_i$ vector) is
\begin{equation}
H_1^{N-1} \left( H_2 (...) H_N \right)^\dagger + c.p. \,, \quad \left( N-1, -1,(...), -1 \right) \,.
\label{eq:leadingN}
\end{equation}
It appears at order $2(N-1)$. If we try e.g. a single non-zero phase $\alpha_1=l 2 \pi/N$, to achieve $A_i =\pi$ requires $l=N/2$ cancelling the $N$ dependence and we obtain $\alpha_1=\pi$ for any $N$.

Other phase-dependent invariants with $a_i$ vector of the type $(N-2, N-2, -2,(...),-2)$ appear at higher order $4(N-2)$.
These can in principle lead to different multiples of $2 \pi/N$, namely single non-zero phases $\alpha_1= \pm \pi/2$ solve $A_i=\pi$ (also no $N$ dependence) - but unnatural coefficient hierarchies are needed for these invariants to dominate.

There are also invariants unique to even $N$, that appear at order $2 (N/2)^2$ (for $N=4$, $2 (N/2)^2 = 4(N-2)$) with $a_i$ vector of type $(N/2,-N/2,(...),N/2,-N/2)$.
These lead to phases with multiples of $2 \pi/N$ without a cancellation of the $N$ dependence, namely for these invariants single non-vanishing phases $\alpha_1=\pm 2\pi/N$ and odd multiples $\alpha_1= \pm l (2\pi/N)$ (for $l<N/2$ and odd) solve $A_i=\pi$. However, as the order of these special invariants grows quadratically with $N$, for these to dominate the coefficient hierarchies would have to be particularly unnatural.

A group generated by $c_N$, $d_n$ and a generator $s_N$ of odd permutations can lead to the same results: the phases are the same and the invariants saturate the permutations of components - e.g. $s_4$ would join the invariants with $a_i$ vectors $(4,-4,0,0)$ and $(4,0,-4,0)$ into the same invariant. The group theoretical details will of course be different with the group having a larger order and different representations - which can be relevant when going beyond the scalar sector, as in the case with $\Delta(27)$ and $\Delta(54)$ \cite{deMedeirosVarzielas:2011zw}.

\subsection{Odd number of scalars \label{sub:odd}}

\begin{figure}
	\centering
		\includegraphics[width=5 cm,keepaspectratio=true]{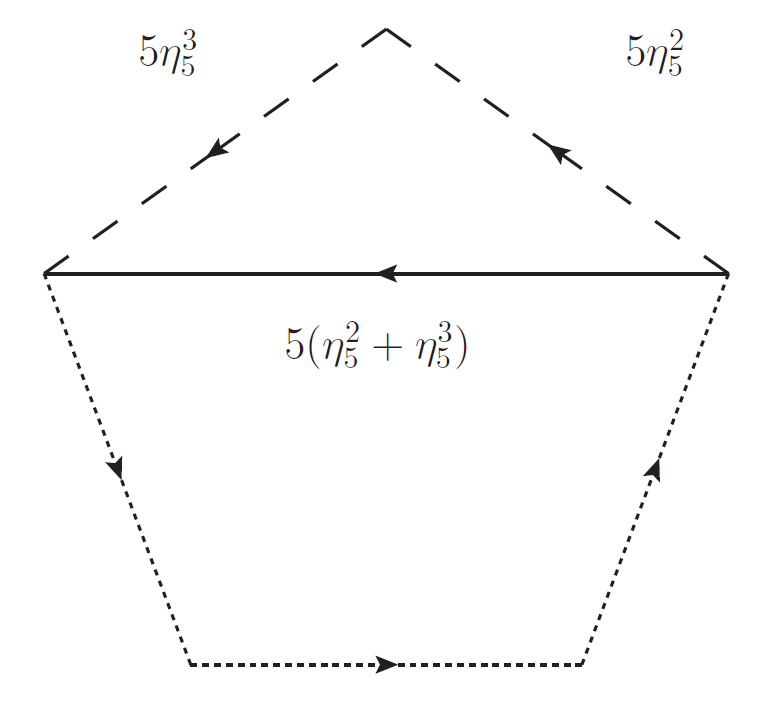}
	\caption{\label{fig:pentagon} The regular pentagon shape associated with minimising solution for $V_5$.}
\end{figure}

We consider now $N=5$ scalars $H_i$ in a single irreducible representation of a group with the cyclic permutation $C_N$. We assume a VEV of type $\langle H_i \rangle = v (e^{\text{i} \alpha_i})$. The phase-dependent invariants have $A_i$ appearing in the quantity
\begin{equation}
V_5 =  (e^{\text{i} A_1} + e^{\text{i} A_2} + e^{\text{i} A_3}+e^{\text{i} A_4}+e^{\text{i} A_5}) + h.c. \,,
\end{equation}
which can get to $10$, but not to $-10$: the number of scalars is odd and $A_i = \pi$ would violate $\sum A_i=0$. The actual minimum that can be obtained is $5(\eta_5^2+\eta_5^3)$
with e.g. $A_i = \pm 2 (2 \pi/5)$.
The minimising $A_i$ manifestly depend on $N=5$, which enables the prospect of new GCPV candidates.
The shape of these minimising solutions in the complex plane can be seen in figure \ref{fig:pentagon} to be made up from two adjacent sides a regular pentagon. 

We consider groups generated by non-commuting $C_5$ generators and a 5-dimensional representation that is a simple generalisation from previously considered cases, with generators $c_5$ (cyclic) and $d_5$ (unit determinant and different powers of $\eta_5$ along the diagonal entries). The order of this particular semi-direct product is $3125$, from $5 n^4$ with $n=5$. This is quite large and fortunately we do not need the group theoretical details. We can verify that at order 8 we have the invariant $H_1^4 (H_2 H_3 H_4 H_5)^\dagger + c.p.$ as a direct consequence of the group treating powers of $4$ as equivalent to powers of $-1$. The associated $a_i$ vector is $(4,-1,-1,-1,-1)$ and the solution $A_i = \pm 2 (2 \pi/5)$ can be obtained with single non-zero $\alpha_1 = \mp 2  (2\pi/5)$. We have therefore found GCPV candidates with VEVs such as $v (\eta_5^{\mp 2} , 1 ,1, 1,1)$ that are a natural result of invariance under $c_5$ and $d_5$.

Higher order invariants are present, as usual with $a_i$ vectors $k(4, -1,-1,-1,-1)$ that are multiples of the lowest order and by saturating $n=5$ with $a_i$ vectors of type $(5, -5,0,0,0)$ and their own multiples.
Another possibility comes from $a_i$ vectors of the type $(3, 3, -2, -2, -2)$ (appearing first at order 12). In order to minimise the $V_5$ of such an invariant with a positive coefficient its $A_i= \pm 2 (2 \pi/5)$. But its phase dependence is different and it is not possible to achieve this with the previous GCPV candidates - it would instead be possible to achieve them e.g. with single non-zero $\alpha_1 =  \mp 2\pi/5$. In order to obtain these phases, the respective phase dependences need to be dominant and to have this occur the coefficients would have to be rather unnatural.

Generalising to any $N$ odd it is possible to generate GCPV candidates with specific integer multiples of $2 \pi/N$.
The difference with even $N$ appears due to $\sum A_i = 0$: if $N/2$ is not integer the $V_N$ minimising conditions depend on $N$ (and have an associated shape in the complex plane that can be seen as two adjacent sides of the regular polygon with $N$ sides). The minimising conditions for $A_i$ are:
\begin{equation}
A_i= \pm \frac{N-1}{2} \frac{2 \pi}{N} \,.
\label{eq:oddNmin}
\end{equation}

The lowest order phase-dependent invariant is given by eq.(\ref{eq:leadingN}). We can solve eq.(\ref{eq:oddNmin}) with single non-vanishing phases:
\begin{equation}
\alpha_1 = \mp \frac{N-1}{2} \frac{2 \pi}{N} \,.
\label{eq:oddNalpha}
\end{equation}
We have therefore distinct GCPV candidates for each $N$ that are a natural result of invariance under $c_N$ and $d_N$. These must violate CP: consider the complex VEV
\begin{equation}
 v \left(\eta_N^{\mp(N-1)/2}, 1,(...),1 \right) \,.
\label{eq:oddNVEV}
\end{equation}
It is CP conserving if there is a transformation $G$ relating the VEV and its conjugate:
\begin{equation}
 G \,. \left(\eta_N^{\pm(N-1)/2}, 1,(...),1 \right) = \left(\eta_N^{\mp(N-1)/2}, 1,(...),1 \right) \,,
\end{equation}
and $G$ leaves the potential invariant. This is not the case as $G_1 = \eta_N^{\pm1}$ ($G$ is diagonal) which explicitly violates this requirement with the invariant in eq.(\ref{eq:leadingN}).

The next unrelated and unsaturated phase-dependent invariants show up at order $4(N-2)$, and are of type $(N-2, N-2, -2,(...), -2)$.
As shown for $N=5$, if this type of invariants dominates (which is unnatural) you need to solve eq.(\ref{eq:oddNmin}) for different $A_i$ and one obtains solutions with a single non-vanishing phase $\alpha_1 = \mp \frac{N-1}{2} \frac{2 \pi}{2N}$.

As in the other cases discussed here and in \cite{deMedeirosVarzielas:2011zw,Varzielas:2012nn}, adding to the group a generator of odd permutations $s_N$ preserves this type of invariants and the associated solutions.

\section{Unmatched phases}

The methods discussed in section \ref{sub:odd} allow GCPV VEVs with powers of $\eta_N$ for odd $N$. This requires $N$ scalars and a discrete group with order increasing very steeply with $N$. We also found interesting solutions from higher orders, but those would require unnatural coefficient hierarchies which are not compatible with the concept of calculable phases.
We aim to address all these drawbacks and enable more GCPV VEVs by exploring cases with unmatched phases, i.e. number of scalars $N$ associated with $c_N$ different from the integer involved in the generator $d_n$: $N \neq n$.


We first reconsider the simpler case of $N=3$ scalars. As pointed out in \cite{deMedeirosVarzielas:2011zw} using $\Delta(3 n^2)$ with $n=6$ ($\Delta(108)$) allows one to choose a representation that results in the same scalar potential as $n=3$ - namely the $3_{02}$ triplet of $n=6$ is effectively the $3_{01}$ triplet of $n=3$, as $d_6$ matrix with powers $\eta_6^0, \eta_6^2, \eta_6^4$ is the same as $d_3$ matrix with powers $\eta_3^0, \eta_3^1, \eta_3^2$.
Therefore the same GCPV candidates can be obtained by the higher order group.

We are interested in new GCPV VEVs, so we explore instead inequivalent representations choices. For example, if $H_i$ is a $3_{01}$ triplet in $n=6$, $H_1^2 (H_2 H_3)^\dagger + c.p.$ is no longer invariant as in general $-1$ is no longer equivalent to $2$ under a $d_n$ transformation. But the higher order combination $\left( H_1^2 (H_2 H_3)^\dagger \right)^2 + c.p.$ remains invariant as $4$ is still equivalent to $-2$. With the lower order combination forbidden by the group and choice of representation, the now lowest order invariant allows new GCPV solutions. $V_3$ is still minimised by $A_i= \pm 2 \pi/3$ but as the invariant for $n=6$ is $2 (2, -1,-1)$ the respective $A_i$ are doubled with respect to the $n=3$ case, and thus the solution with only non-vanishing $\alpha_1= \mp 2\pi/6$ appears. We can then have VEVs of type $v(\eta_6^{\mp},1,1)$ which are good GCPV candidates as the potential does not remain invariant under the transformation that relates them to their conjugates.

Note that $a_i$ vectors such as $(5,-1,-1)$ for $n=6$ correspond to combinations that are not invariant because they don't obey $\sum a_i = 0$. This restriction also forbids phase-dependent invariants from cases with $N=3$ and $n$ that are not multiples of 3, with the sole exception of those that saturate $n$ like $(n,-n,0)$.
In section \ref{sub:oddre} we will discuss this type of case in more detail, but we note that with invariant $(n,-n,0)$ the $V_3$ minimising solutions $A_i= \pm 2 \pi/3$ can be obtained with single non-vanishing phases $\alpha_1 = \pm l (2\pi/3n)$ for any $l<3n/2$ that is not a multiple of $3$ (those lead instead to $A_i=0$). We have then $2n$ GCPV candidates for each $n$ that is not a multiple of 3 that are a natural consequence of generators $c_3$ and $d_n$. For $n=4$ we have $\alpha_1= l (\pm 2\pi/12)$, for $l=1,2,4,5$, some obtained previously although with this method we may be obtain them naturally from a smaller group (e.g. $l=2$).


\subsection{Even number of scalars revisited \label{sub:evenre}}

We now attempt to obtain new phases from $N=4$ by going to $n=8$. We avoid representations that effectively reduce to $n=4$ in order to disable the lowest order combination $(3,-1,-1,-1)$, while preserving $2 (3, -1, -1, -1)$. With $A_i = \pi$ this leads to e.g. single non-vanishing $\alpha_1 = \pm \pi/2$ without requiring unnatural hierarchies, so VEVs of type $v(\eta_4,1,1,1)$ become natural.
By the same reasoning, invariants of type $2 (2,2,-2,-2)$ can lead to phases of $\pi/4$, but require unnatural hierarchies to dominate over $2(3,-1,-1,-1)$ for $n=8$. Instead with $n=16$ we obtain them naturally through the now leading $4(3,-1,-1,-1)$ invariant.

We can generalise to any number of scalars $N$ with $c_N$ and $d_n$ with $n=k N$ containing different powers of $\eta_{kN}$ chosen so that $d_n$ does not take the form of $d_N$. Then the lowest order phase-dependent invariant is:
\begin{equation}
\left( H_1^{N-1} \left( H_2 (...) H_N \right)^\dagger \right)^k + c.p. \,, \quad k \left( N-1, -1,(...), -1 \right) \,.
\label{eq:leadingkN}
\end{equation}
With positive coefficient, we know that such an invariant for $k=1$ achieves $A_i = \pi$ with a single non-vanishing phase $\alpha_1=\pi$. For other $k$ where the $a_i$ vector is comparably multiplied by $k$, single non-vanishing phase $\alpha_1=\pm \pi/k$ must be solutions. But we may also obtain additional solutions $\alpha_1= (\pm l\pi/k)$ for any odd $l$ with $l \leq k$, for a total of $k$ solutions (this total includes the phase $\pi$ for $k=l$ if $k$ is odd).
These are suboptimal GCPV candidates:
$G_1 \eta_{2k}^{\mp l} = \eta_{2k}^{\pm l}$, $G_1=\eta_{2k}^{\pm 2l}$. The invariants involve multiples of $k$ so the potential is invariant under $G$.

More interesting methods arise in $N=4$ and $n=6$, or more generally for any even scalars $N$ and any even $n$ not a multiple of $N$.
As $N$ and $n$ are both even it is always possible to have invariants of type $(n/2,-n/2,(...),n/2,-n/2)$ appearing at order $N (n/2)$. These invariants are similar to the subleading invariants that have appeared for even $N=n$, $(N/2,-N/2,(...),N/2,-N/2)$, so it is not surprising that through them $A_i=\pi$ can be obtained from single non-vanishing phases $\alpha_1= \pm 2\pi/n$ and odd multiples $\alpha_1= \pm l 2\pi/n$ ($l<n/2$ and odd). The choice of $n$ and the $\sum a_i$ requirement combine to eliminate the otherwise dominant combinations $k(n-1,-1,(...),-1)$ as $n$ is not a multiple of $N$ (e.g. $(5,-1,-1,-1)$ for $N=4$ and $n=6$ is not invariant).
There are other phase-dependent invariants $(n,-n,0,(...),0)$ appearing at order $2n$, but as long as the sign of the coefficient of their invariants is negative one can always satisfy $A_i=0$ for them with phases that are integer multiples of $2 \pi/n$. This type of solution can thus be obtained without requiring unnatural hierarchies.
While this method obtains natural GCPV candidates of type $v(\eta_n^{\pm l},1,(...),1)$ for even $n$ and odd $l$, they are suboptimal as the potential only has powers of $\pm n/2$ and $\pm n$ and remains invariant under the transformation that relates them to their conjugates:
$G_1 \eta_n^{-l} = \eta_n^l$, $G_1=\eta_n^{2l}$.

If instead there is a negative coefficient for the dominating invariant
\begin{equation}
\left( H_1 H_2^\dagger \right)^n + c.p. \,, \quad (n,-n,0,(...),0) \,,
\label{eq:leadingn}
\end{equation}
the solution $A_i = \pi$ can be obtained from single non-vanishing phases $\alpha_1= \pm l \pi/n$, for $l<n$ and odd. This applies also if $n$ is odd, where the only invariants are of the type eq.(\ref{eq:leadingn}). We have natural GCPV candidates such as
\begin{equation}
v \left( \eta_{2n}^{\pm l},1,(...),1 \right) \,, \quad \text{for odd} \quad l < n \,.
\label{eq:evennVEV}
\end{equation}
If we had positive coefficient instead the phases want to achieve $A_i=0$ and we get candidates that are just like eq.(\ref{eq:evennVEV}) but with even $l<n$.
In any case these candidates are suboptimal for $n$ odd:
$G_1 \eta_{2n}^{\mp l} = \eta_{2n}^{\pm l}$, $G_1=\eta_{2n}^{\pm 2l}$ and the potential has powers of $\pm n$, with $\eta_{2n}^{\pm 2nl}=1$ so $G$ leaves the potential invariant. This is consistent with previous results, as the odd $n$ case here is effectively equivalent to the case $2n$ with invariant $((2n)/2,-(2n)/2,(...),(2n)/2,-(2n)/2)$ considered above - for $n$ odd the symmetry group is smaller but the potential and candidate solutions are the same.
For even $n$ the natural GCPV candidates in eq.(\ref{eq:evennVEV}) with $l$ odd must violate CP, as the additional powers of $\pm n/2$ present in the potential make it not invariant under $G$, with residual $\eta_{2n}^{nl}$ (the $l$ even candidates obtained from $A_i = 0$ are therefore also suboptimal).

As a special case, we consider $N=2$. The optimal candidates that must violate CP appear only with the last method and for $n$ even, and for $N=2$ you always have $n=kN$ where the respective $G$ was a symmetry of the potential. Indeed, this was noted to be case explicitly in \cite{Branco:1983tn} for $k=2$, where the candidate is $(\eta_4,1) = (\text{i},1)$. We can confirm this applies for any $k$: the candidate VEVs are of type $(\eta_{2k}^{\pm l},1)$, $G_1 \eta_{2k}^{\mp l} = \eta_{2k}^{\pm l}$ implies $G_1=\eta_{2k}^{2l}$ and the potential only has invariants multiple of $(k,-k)$. With this analysis we have generalised to non-renormalisable potentials the conclusion of \cite{Branco:1983tn} that more than 2 scalars are required for GCPV VEVs that must violate CP.

\subsection{Odd number of scalars revisited \label{sub:oddre}}

For any odd $N$ we take $n=k N$. By selecting the representation appropriately, $d_n$ does not reduce to $d_N$, disallowing the invariant from eq.(\ref{eq:leadingN}) but preserving the one in eq.(\ref{eq:leadingkN}) as the now lowest order phase dependence.
To minimise $V_N$, eq.(\ref{eq:oddNmin}) applies.
For $k=1$ we have single non-vanishing phases given by eq.(\ref{eq:oddNalpha}), therefore single non-vanishing phases $\alpha_1 = \mp \frac{N-1}{2} \frac{2 \pi}{kN}$ must be new solutions. For $n=2$ this leads to the previously known solutions $\alpha_1 = \mp\frac{N-1}{2} \frac{2 \pi}{2N}$ without requiring unnatural coefficient hierarchies.
In fact we have $2k$ solutions in total:
\begin{equation}
v \left( \eta_{kN}^{\mp \frac{(2l-1)N-1}{2}},1,(...),1 \right) \quad \text{for} \, l \leq k  \,.
\label{eq:oddNkVEV}
\end{equation}
These are good GCPV candidates as $G_1=\eta_{kN}^{\mp ((2l-1)N-1)}$, so in general $G$ does not leave the potential invariant.

We turn to cases with $n$ not a multiple of $N$. We start with the case where $n$ and $N$ share a prime factor as a generalisation to odd $N$ of what occurs with even $N$ and $n$. With $p$ the smallest shared prime factor, $N=l p$ and $n=m p$, it is possible to obtain an invariant by dividing the $N$ scalars into $p$ groups of size $l$ and distributing the phases as if you had $N=p$ instead: $m(p-1)$ to one of the $p$ groups and $-m$ to the remaining $p-1$ groups.
These appear at order $2 m l (p-1)$.
E.g. for $N=9$, $n=6$ ($p=3$), we divide into $3$ groups with $4$ and $-2$: $(4,4,4,-2,(...),-2)$ (corresponding to the $N=3$, $n=6$ invariant $(4,-2,-2)$).
The $V_N$ minimising solutions in eq.(\ref{eq:oddNmin})  could be solved with e.g. only non-vanishing phases $\alpha_1 = \mp \frac{(l p-1 ) \pi}{m l p}$.
But the invariants in eq.(\ref{eq:leadingn}) appear at order $2 m p$ which is always lower order for $p>2$, as then $l < p/(p-1)$ only for $l=1$.
Further, the phases obtained are not integer multiples of $2\pi/n$ so they are generally not preserved by the eq.(\ref{eq:leadingn}) invariants. Therefore to obtain these solutions with this method requires unnatural coefficient hierarchies, which makes it less promising despite the relatively small $n$ used. Instead one can obtain this type of phases with the other method by keeping the same $N = l p$ and replacing $n=m p$ with $n = mN$: the group has $m(N-1,-1,(...),-1)$ at lower order than $(mN,-mN,0,(...),0)$.

Finally, if $n$ and $N$ do not share any prime factor the only type of invariants are those of eq.(\ref{eq:leadingn}). This is a consequence of $\sum a_i = 0$: we attempt to fill the $a_i$ vector by assigning $(n-m)$ any $r$ times, and $-m$ the remaining $N-r$ times. $\sum a_i = r n- m N$ vanishes for $r=mN/n$ which can only be an integer if $N$ and $n$ share a prime factor.
For dominating eq.(\ref{eq:leadingn}) invariants with a positive coefficient, $A_i=0$ can be obtained from $v\left( \eta_{n}^{\pm l},1,(...),1 \right)$ and $l \leq n$ - not all of these candidates with powers of $\eta_n$ were obtainable directly by other methods. For $G_1 \eta_{n}^{\mp l} = \eta_{n}^{\pm l}$, $G_1=\eta_{n}^{\pm 2l}$, so in general $G$ leaves the potential invariant.
With a negative coefficient though, the solutions in  eq.(\ref{eq:oddNmin}) can be obtained with $2n$ GCPV candidates
\begin{equation}
v \left( \eta_{nN}^{\mp \frac{(2l-1)N-1}{2}},1,(...),1 \right) \,, \quad \text{for} \, l \leq n  \,.
\label{eq:oddnVEV}
\end{equation}
For these, $G_1 = \eta_{nN}^{\pm ((2l-1)N-1)}$, so $G$ does not leave the potential invariant. 
 

\section{Conclusion}

We have presented several methods that lead to calculable phases for $N$ scalars placed in single irreducible representations of discrete non-Abelian groups that contain a generator of cyclic permutations, $c_N$ that does not commute with another other cyclic generator $d_n$ (and optionally a generator of odd permutations $s_N$).

Some of the complex vacuum expectation values are CP conserving in the absence of other fields, due to accidental symmetries under which the scalar potential is invariant. We showed this always occurs for $N=2$.
These may become CP violating as long as the Lagrangian contains other invariants that eliminate the accidental symmetries.

We have also obtained optimal candidates for geometrical CP violation with complex vacuum expectation values that must violate CP:
eqs.(\ref{eq:oddNVEV}),
(\ref{eq:oddNkVEV}),
(\ref{eq:oddnVEV}),
and additionally eq.(\ref{eq:evennVEV}) for even $n$. These and the other analysed cases are summarised in table \ref{ta:1}.

\begin{table}[h]
	\centering
\begin{tabular}{|c|c|c|}
\hline
Case & $a_i$ vector & Phase, eq. if optimal \\ \hline
Even $N$ &  $(N-1, -1, (...), -1)$ & $\pi/2$ \\
Even $N$ & $(N/2,-N/2,(...),N/2,-N/2)$ & $\pm l \frac{2 \pi}{N}$ ($l$ odd) \\
$n=kN$ & $k(N-1, -1, (...), -1)$ & $\pm l \frac{\pi}{k}$ ($l$ odd) \\
$n\neq kN$ & $(n, -n, 0, (...), 0)$ & $\pm l \frac{2\pi}{2N}$, eq.(\ref{eq:evennVEV})$^{(a)}$ \\ \hline
Odd $N$ & $(N-1, -1, (...), -1)$ &	$\mp \frac{N-1}{2} \frac{2 \pi}{N}$, eq.(\ref{eq:oddNVEV}) \\ 
$n=kN$ & $k(N-1, -1, (...), -1)$ &	$\mp \frac{(2l-1)N-1}{2} \frac{2 \pi}{kN}$, eq.(\ref{eq:oddNkVEV}) \\
$N=lp$, $n=mp$, prime $p$	& $m([p-1]_l,-1,(...),-1)$ &	$\mp \frac{(l p-1 ) \pi}{m l p}$  \\
$N\neq kp$, $n=mp$, prime $p$ & $(n, -n, 0, (...), 0)$ &	$\mp \frac{(2l-1)N-1}{2} \frac{2 \pi}{nN}$, eq.(\ref{eq:oddnVEV}) \\ \hline
		\end{tabular}
		\caption{The four even $N$ cases at the top have $A_i=\pi$, the four odd $N$ cases at the bottom have $A_i=\pm \frac{N-1}{2} \frac{2 \pi}{N}$, eq.(\ref{eq:oddNmin}). $^{(a)}$ denotes that the candidates in eq.(\ref{eq:evennVEV}) are only optimal when $n$ is even and additionally $l$ is odd. $[p-1]_l$ represents a block of entries $(p-1)$ repeated $l$ times. \label{ta:1}}
\end{table}

The scalars need not be Standard Model doublets - as long as there is some mechanism that justifies $\sum a_i=0$ our results apply.
This type of formalism, invariants and geometrical CP violating candidates may have applications within family symmetry models, where it would be particularly interesting to consider the possibilities for fermion masses and mixing as one can expect very specific predictions for the CP violating quantities obtained.

\acknowledgments

This work was supported by DFG grant PA 803/6-1 and partially supported through the FCT by project PTDC/FIS/098188/2008.

\bibliography{refs}

\end{document}